\documentclass{article}
\usepackage{graphicx}
\usepackage{epsfig}
\usepackage{amsfonts}
\usepackage{amsmath}
\usepackage{color}
 \usepackage{amssymb}
\date{\today}

\newcommand{\insertplot}[5]{\begin{figure}
 \hfill\hbox to 0.05in{\vbox to #5in{\vfill
 \inputplot{#1}{#4}{#5}}\hfill}
 \hfill\vspace{-.1in}
 \caption{#2}\label{#3}
 \end{figure}}
 \newcommand{\inputplot}[3]{
 \special{ps: plotfile #1}
}
\newcounter{fig}

\newcommand{\ee}{\end{equation}}
\newcommand{\eea}{\end{eqnarray}}
\newcommand{\be}{\begin{equation}}
\newcommand{\bea}{\begin{eqnarray}}

\newcommand{\e}{\mbox{e}}

\newcommand{\R}{{\rm I \hspace{-0.52ex} R}}

\begin{document}

 \title{Q-balls and charged Q-balls in a two-scalar field theory with generalized Henon-Heiles potential
}

\author{
{\large Y. Brihaye}$^{1}$,
{\large F. Buisseret}$^{2,3}$
\\
$^1$ Service de Physique de l'Univers, Champs et Gravitation, \\ Universit\'{e} de Mons, UMONS  Research Institute for Complex Systems,\\  20 Place du Parc, 7000 Mons, Belgium.\\
$^2$ Service de Physique Nucl\'{e}aire et Subnucl\'{e}aire, \\ Universit\'{e} de Mons, UMONS Research Institute for Complex Systems,\\ 20 Place du Parc, 7000 Mons, Belgium. \\ 
$^3$ CeREF , Chaussée de Binche 159, 7000 Mons, Belgium. \\
}
\maketitle
\begin{abstract}
We construct Q-ball solutions from a model consisting of one massive scalar field $\xi$ and one massive complex scalar field $\phi$ interacting via the cubic couplings $g_1 \xi \phi^{*} \phi + g_2 \xi^3$, typical of Henon-Heiles-like potentials. Although being formally simple, these couplings allow for Q-balls. In one spatial dimension, analytical solutions exist, either with vanishing or non vanishing $\phi$. In three spatial dimensions, we numerically build Q-ball solutions and investigate their behaviours when changing the relatives values of $g_1$ and $g_2$. For $g_1<g_2$, two Q-balls with the same frequency exist, while $\omega=0$ can be reached when $g_1>g_2$. We then extend the former solutions by gauging the U(1)-symmetry of $\phi$ and show that charged Q-balls exist.
 \end{abstract}

\section{Introduction}\label{sec:intro}
Scalar fields appear in numerous sectors of physics as solid state physics, elementary particles, supersymmetry,
gravity and cosmology. On the theoretical side, numerous classical solutions have been obtained in field theories involving scalar fields. Among them are soliton-like solutions, and perhaps the most popular ones are Q-balls, which are nontopological solitons appearing in classical field theories presenting  a global symmetry \cite{Coleman:1985ki,Lee:1991ax}. A simple example involves just one complex scalar field, $\phi$, with a  potential invariant under the $U(1)$ phase group and obeying some conditions derived in \cite{Coleman:1985ki}. In this pioneering work it is demonstrated that the existence of Q-balls needs high-degree interacting terms, e.g. at least $(\phi^*\phi)^2$, which are not compatible with renormalisation. The standard  Q-ball in three spatial dimensions is spherically symmetric and is characterized by the ansatz $\phi(r) = {\rm exp}(i \omega t) f(r)$ with a constant  harmonic frequency $\omega$  and a real, radial, function $f(r)$ which can present zero or a finite number of nodes. The fundamental solution has no node and decreases  monotonically from a finite value at the center, 
  say $f(0)$, to zero at spatial infinity. Node solutions are interpreted as excitations of the fundamental solution. Enlarging the ansatz to an axial symmetry allows for other types of Q-balls, 
	namely spinning solutions \cite{Volkov:2002aj}.The non-linear character  of the  field equations requires to solve them by approximation or numerical techniques, even when the interaction is represented by a polynomial in the squared modulus $\phi^* \phi$.

The minimal theory involving a single self-interacting complex scalar field can be enlarged in several directions, namely by coupling to electromagnetism \cite{charged_qb} or considering two or 
more extra complex fields \cite{Brihaye:2007tn}. One of the simplest way along the latter direction consists in supplementing the minimal theory by one real massive scalar field $\xi$. For example, the Friedberg-Lee-Sirlin model \cite{sirlin} provides an interesting case of a renormalizable two-component scalar field theory with natural interaction terms of degree four in the two fields: ${\cal L}= \partial_{\mu}\phi^{*} \partial^{\mu}\phi  + 
 \frac{1}{2}\partial_{\mu}\xi \partial^{\mu}\xi - d^2 \xi^2  \phi^{*} \phi + \frac{g^2}{8} (\xi^2-\xi^2_{vac})^2$ with $d$ and $g$ coupling constants. In this model, the complex scalar becomes massive due to the coupling with the real scalar
 field, since the latter has a finite vacuum expectation value $\xi_{vac}$ generated via a symmetry-breaking
 potential. In this case both the harmonic time-dependence of the complex scalar and its coupling with the
 real field allow for Q-balls to exist. Extensions gauging the U(1) symmetry of this model have been emphasized, see e.g. \cite{Loiko:2019gwk} and references therein.

 In this paper, we will focus on a model involving a real and complex scalar field proposed in \cite{Nugaev:2014ima, Nugaev:2019vru}:
 \be
 S = \int d^{D+1} x  \bigg[ 
 \partial_{\mu}\phi^{*} \partial^{\mu}\phi - m^2 \phi^{*} \phi  + 
 \frac{1}{2}\partial_{\mu}\xi \partial^{\mu}\xi - \frac{1}{2} M^2 \xi^2 + g_1 \xi \phi^{*} \phi + g_2 \xi^3
 \bigg]
 \label{lagrangian}
 \ee
 with Minkowski metric $\eta=(+---)$ and where $g_1$, $g_2$ are positive real coupling constants. In contrast to \cite{sirlin}, the masses of the two fields are set by hand and the polynomial interaction is cubic in the two scalar fields. This interaction term is inspired by the celebrated Heinon-Heiles potential in classical mechanics \cite{hh}. 
 
 In \cite{Nugaev:2014ima}, the model (\ref{lagrangian}) was studied in one spatial dimension and an explicit solution was obtained for $g_2=g_1$. In this work the cases $D=1$ and $D=3$ will be studied in Secs. \ref{sec:D1} and \ref{sec:D3} respectively, without restriction on $g_2$ and $g_1$, after having defined our ansatz in Sec. \ref{sec:ans}. Finally, the U(1) symmetry of the model will be gauged, and charged Q-balls will be built in Sec. \ref{sec:charged}.

\section{Q-ball ansatz}\label{sec:ans}

We are interested in classical solutions associated with the model (\ref{lagrangian}). Noting $r$ ($z$) the $D$-dimensional radial variable in $D>1$ ($D=1$), we make a non-rotating Q-ball ansatz for $\phi$ and also ask for a radial form for $\xi$:
\be
    \phi = e^{i \omega t} \frac{F(r)}{\sqrt 2} \ \ \ , \ \ \ \xi = G(r),
		\label{ansatz}
\ee
$F$ and $G$ being real functions. The corresponding equations of motion take the form
\begin{subequations}\label{eom0}
\begin{eqnarray}
        F'' + \frac{D-1}{r} F'& =& \Omega^2\, F - g_1 F G\ , \label{eom1}\\
				G'' + \frac{D-1}{r} G' &=& M^2\, G - \frac{g_1}{2} F^2 - 3 g_2 G^2  \label{eom2} \ , 
\end{eqnarray}

with 
\begin{equation}\label{Omegadef}
	\Omega^2 = m^2 - \omega^2.
\end{equation}	
\end{subequations}

For the regular solutions of Q-ball-type that we are interested in,
the non-linear system above has to  be solved with the boundary conditions
\be\label{bound}
F'(0) = 0 \ , \ G'(0) = 0 \ , \ G(0) = C \ , \ F(\infty) = 0 \ , \ G(\infty) = 0 
\ee
while the frequency $\omega$ has to be fine-tuned as function of the central value $C$. It is convenient to  use $C$ as a control parameter. The case $D=1$ deserves a separate study since analytical solutions exist, see next section. Although the case $D=3$ is physically motivated, the field equations do not admit (up to our knowledge)
analytical solutions, but the equations can be treated by numerical methods. We integrate the equations numerically by using the solver
COLSYS \cite{colsys}. The axis of radial coordinate was discretized by about 400 points and the solutions were obtained with an error less than $10^{-8}$.

The Q-ball solutions can be characterized by several physical
quantities, namely their energy $\tilde E$ and conserved Noether charge $\tilde Q_N$. They are given respectively
by the integrals
\be
         \tilde E = \int_{\R^D} d^D x \  T_0^0  \ \  , \ \ \tilde Q_N = \int_{\R^D} d^D x \ J^0 ,
\ee
 where the definition of $T_0^0$ is standard and where $J^0 = i \left( \phi \frac{\partial}{\partial t} \phi^* - \phi^* \frac{\partial}{\partial t} \phi\right)$. With the ansatz above, $ \tilde E \equiv   V_{D-1} E$ and $\tilde Q_N \equiv   V_{D-1} Q$, where $V_{D-1}$ is the volume of the $(D-1)$-sphere ($V_0=1 \ , V_2 = 4 \pi$) and where $E$ and $Q$ are evaluated by simple integrals:
\be\label{energy}
E = \int_0^{\infty}dr\ r^{D-1}
	\Bigl( \frac{1}{2} (F'^2 + (m^2 + \omega^2) F^2 + G'^2 + M^2 G^2) - \frac{g_1}{2} F^2 G - g_2 G^3 \Bigr),
\ee
\be
Q = \omega \int_0^{\infty}dr\ r^{D-1} F^2.
\ee

\section{$D=1$: Henon-Heiles effective Hamiltonian}\label{sec:D1}
\subsection{Effective potential}
For $D=1$, the equations of motion are equivalent to the equations of motion of the Henon-Heiles-type Hamiltonian
\begin{equation}\label{hhham}
	H=\frac{1}{2}\left(P^2_F+P^2_G-\Omega^2 F^2-M^2 G^2\right)+\frac{g_1}{2} F^2 G+g_2 G^3,
\end{equation}
with $P_F=F'$ and $P_G=G'$, the derivative being taken with respect to the spatial coordinate $z$, here seen a the temporal parameter of the effective Hamiltonian. 

The solutions of the equations of motion are therefore related to the movement of an effective particle in the potential
\begin{equation}\label{effpot}
	V(F,G)=-\frac{\Omega^2}{2} F^2-\frac{M^2}{2} G^2+\frac{g_1}{2} F^2 G+g_2 G^3,
\end{equation}
that is
\begin{equation}
	F''=-\partial_F V,\quad G''=-\partial_G V.
\end{equation}

We note that $V(-F,G)=V(F,G)$ and that $\partial_GV(F,G<0)>0$: no bounded trajectory is expected if $G$ becomes negative; $F$ however may change sign. $V$ has a local maximum in $(F,G)=(0,0)$ for all values of the parameters, and $V(0,0)=0$. If $\frac{g_1}{g_2}\leq 3\frac{\Omega^2}{M^2}$, $V$ has a saddle point in $\left(0,\frac{M^2}{3g_2}\right)$ and no other extremal point. The most favourable situation for the existence of nontrivial solutions is rather 
\begin{equation}\label{cond1}
\frac{g_1}{g_2}> 3\frac{\Omega^2}{M^2} , 
\end{equation} 
where $V$ has a minimum in $\left(0,\frac{M^2}{3g_2}\right)$, the minimal value being $-\frac{M^6}{54g_2^2}$, and where $V$ has also a saddle point in $\left(\frac{\Omega^2}{g_1},\sqrt 2\frac{\Omega}{g_1}\sqrt{M^2-3\frac{g_2}{g_1}\Omega^2}\right)$. 

We remark that the original Henon-Heiles Hamiltonian \cite{hh} would be obtained by setting $M^2=\Omega^2=-1$, $g_1=2$ and $g_2=-1/3$. This case will not be investigated here since soliton-like solutions are rather found for positive $M^2$ and $\Omega^2$, but it has motivated a tremendous number of studies, to which we refer the reader. Many references and original results regarding the existence and types of trajectories in the original Henon-Heiles model can be found in \cite{10.3389/fspas.2022.945236,zlotos}.

\subsection{Explicit solutions}
\begin{figure}
	\centering
	\includegraphics[width=0.4\linewidth]{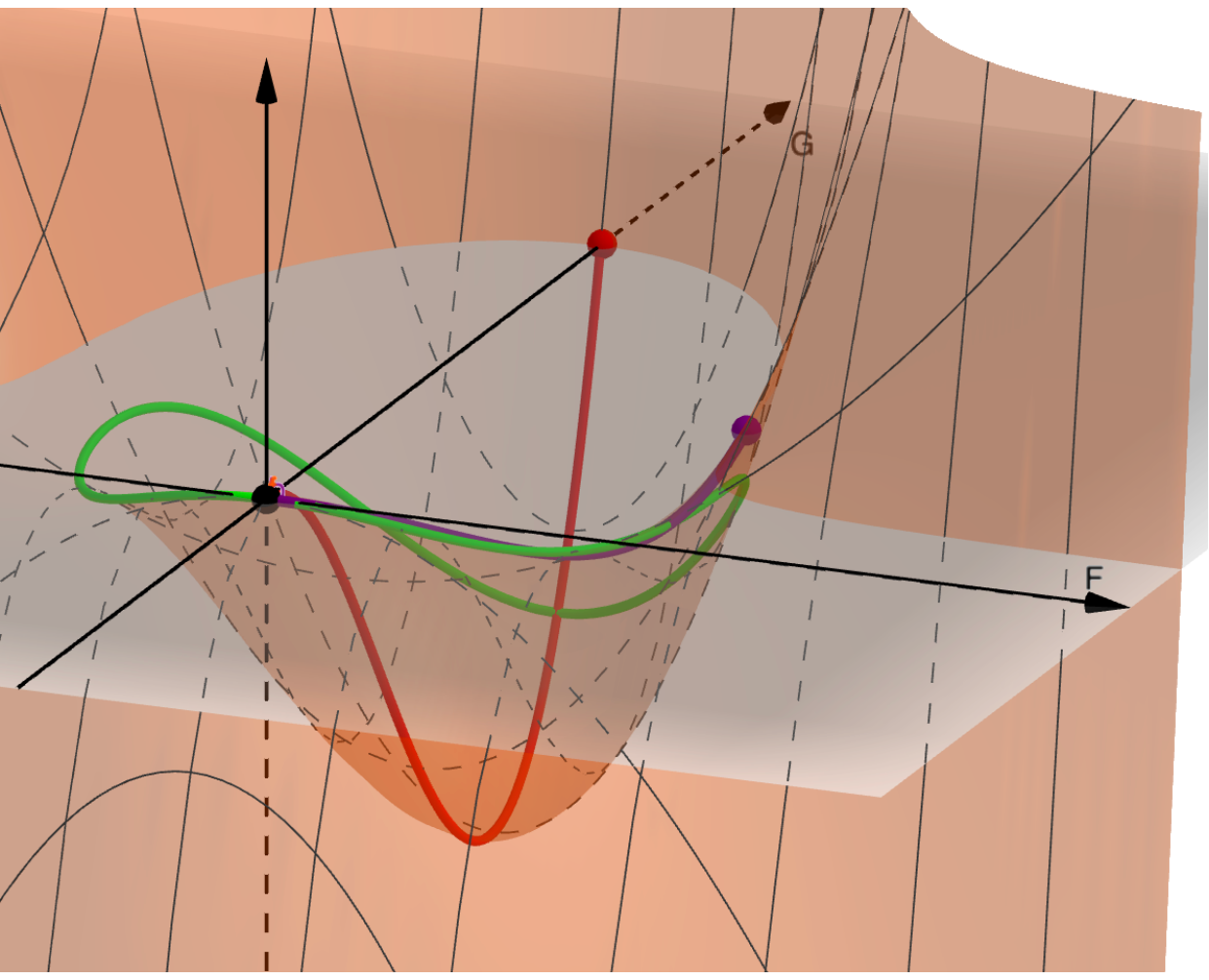}
	\caption{Effective potential (\ref{effpot}) and typical allowed solutions (solid lines). The sketched solutions actually correspond to (\ref{profil1}) (red), (\ref{profil2}) (purple) and (\ref{profil3}) (green). The displayed shape for $V$ has been obtained for $M=2.11166$, $m=1$, $g_1=2$, $g_2=1$, $\Omega=0.8$, i.e. a configuration for which the three solutions may exist. The plot was made using GeoGebra software.}
	\label{fig:potential}
\end{figure}

The profiles of $F(z)$ and $G(z)$ we look for reach their global maxima in $F(0)$ and $G(0)$ respectively. Moreover, they tend to zero as $z\to\pm\infty$: $F(\pm\infty)=G(\pm\infty)=0$ and $(P_F,P_G)=(0,0)$ as $(F,G)=(0,0)$ so the motion of the effective particle has zero total energy. The $V=0$ curve has the equation 
\begin{equation}\label{V0}
	F^2=G^2\frac{M^2-2g_2 G}{g_1G-\Omega^2}
\end{equation}
which, imposing $F^2\geq 0$, is defined for 
\begin{equation}
	G=\left\{0,\frac{\Omega^2}{g_1}<G<\frac{M^2}{2g_2}\right\}.
\end{equation}
As sketched in Fig. \ref{fig:potential}, the soliton starts from $(0,0)$ at $z\to-\infty$, then $(F,G)$ reach the $V=0$ curve at $z=0$, that is 
$F(0)$ and $G(0)$ linked by (\ref{V0}), and finally turn back to $(0,0)$ at $z\to+\infty$. If $G(0)$ reaches the maximal value $\frac{M^2}{2g_2}$, then the only allowed solution is $F=0$. If $G(0)$ approaches the minimal value $\frac{\Omega^2}{g_1}$, then $F(0)$ becomes larger and larger.

Analytical solutions can be pointed out. First, a solution in which $F$ is trivial exists for any nonzero value of the parameters:
\begin{subequations}\label{profil1}
	\begin{eqnarray}
		F(z)&=&0, \\
		G(z)&=&\frac{M^2}{2g_2}\frac{1}{\cosh^2\left(\frac{M}{2}z\right)}.
	\end{eqnarray}
\end{subequations}

Regarding solutions with nonvanishing $F$ and $G$, Hamiltonian (\ref{hhham}) is known to be separable in three specific cases: $6g_2=g_1$ and $\Omega^2=M^2$ (Sawada-Kotera), $g_2=\frac{8}{3}g_1$ and $M^2=16\Omega^2$ (Kaup-Kupershmidt), $g_2=g_1$ and $\Omega^2$, $M^2$ arbitrary (KdV5) \cite{verhoeven}. The corresponding Hamilton-Jacobi equations may be solved in terms of hyperelliptic integrals \cite{verhoeven,WOJCIECHOWSKI1984277}. An other path to build solutions was followed in \cite{conte}, in which it is shown that solitary-wave-type solutions of Riccati equations may be used to build soliton-like solutions of the equations (\ref{eom1}), (\ref{eom2}). 
When $g_2=g_1$, the solution denoted KdV5$_1$ in \cite{conte} and also given in \cite{Nugaev:2014ima} may be quoted:
\begin{subequations}\label{profil2}
	\begin{eqnarray}
		F(z)&=&\frac{2\Omega}{g_1}\frac{\sqrt{M^2-4\Omega^2}}{\cosh(\Omega z)}, \\
		G(z)&=&\frac{2\Omega^2}{g_1}\frac{1}{\cosh^2(\Omega z)}.
	\end{eqnarray}
\end{subequations}
Solution (\ref{profil2}) is well-defined for 
\begin{equation}\label{omegam}
\omega^2\geq m^2-\frac{M^2}{4};
\end{equation}
it reduces to (\ref{profil1}) when the lower bound is reached. Note that parameters such that $M\geq 4m$ may in principle allow for $\omega=0$ solutions.

The shapes of other analytical solutions are listed in Table 1 of \cite{conte}. They correspond to specific values of the ratio $\frac{\Omega^2}{M^2}$ in terms of $g_1$ and $g_2$. For example, we find that the solution called $HH_1$ in \cite{conte} solves our equations of motion for the following explicit values of the parameters:
\begin{subequations}\label{profil3}
	\begin{eqnarray}
		F(z)&=&\frac{6\Omega}{g_1}\sqrt{2\left(3\frac{g_2}{g_1}-1\right)}\ \frac{\sinh(\Omega z)}{\cosh^2(\Omega z)}, \\
		G(z)&=&\frac{6\Omega^2}{g_1}\frac{1}{\cosh^2(\Omega z)},\\
		&& {\rm with\quad} M^2=2\Omega^2\left(9\frac{g_2}{g_1}-1\right),\quad {\rm and\quad} \frac{g_2}{g_1}>\frac{1}{3}.
	\end{eqnarray}
\end{subequations}
This last solution has $P_F(0)\neq 0$, and hence do not reach the $V=0$ curve before turning back to the origin; $G\in\left[0,\frac{6\Omega^2}{g_1}\right]$. Solutions similar to (\ref{profil3}) will not be investigated further here since we consider $\omega$ as unspecified, and a priori independent of the coupling constants. Other solutions may also be constructed numerically, but this task will be performed at $D=3$, the $D=1$ case being seen as a test case to analytically understand the existence and types of solutions.

\section{$D=3$: Q-Balls}\label{sec:D3}
The four physical parameters $m$, $M$, $g_1$, $g_2$ can be redefined by appropriate
rescaling of the radial coordinate and of the two scalar fields. We will use this freedom  to set $m=1$ and $g_2=1$ in the rest of this work so that 
$M$, $g_1$  become the relevant parameters for the study of the solutions.
The numerical analysis of the equations reveals that the spectrum of Q-balls depends significantly on $M$ and $g_1$.

\subsection{A special solution}
First, let us notice that the system (\ref{eom0}) possesses a unique solution with $F(r)=0$ and $G(r)\neq0$  with no node.
Because this solution play an important role in the classification of the solutions 
we find convenient to note it $G_0(r,M)$. It exists irrespectively of $g_1$ but depends on $M$. 

Inspection of the equation (\ref{eom2}) shows that the scaling relation $G_0(r,M) = M^2 G(Mr,1)$ holds 
for any spatial dimension. Accordingly the function $G_0(r,M)$ smoothly approaches the null function
  in the limit $M \to 0$. We could not find a closed form for $G_0(r,M)$. For later use we note that we found the values
	$G_0(0,1) \approx 1.3972$ and $G_0(0,2) = 4 G_0(0,1) \approx 5.5889$.
Remark that the	solution (\ref{profil1}) at the centre is equal to $0.5$ (resp. $2$) for $M=1$ (resp. $M=2$);
 the larger values of $G_0(0)$ for $D=3$ are due to the damping term $\frac{2}{r}G'$ in $D=3$. The energy at $r=0$ has indeed to be larger than 0 for the effective particle to reach $G=0$ at infinity.
 
We notice that profiles such as (\ref{profil3}), with nonzero $F'(0)$ or $G'(0)$, are not allowed in $D=3$: A power expansion around the origin shows that only the boundary conditions (\ref{bound}) may lead to regular solutions at origin. 



\subsection{The case $g_1=g_2$} 

For definiteness, let us first discuss the solutions in the case $g_1=g_2$, owing that the explicit solution (\ref{profil2}) exists for $D=1$ and will be compared to the numerically obtained following solutions. 

It is found that Q-ball solutions exist for a finite interval 
of the frequency $\omega$, that is for $\omega \in [\omega_m, 1.0]$
where $\omega_m$ depends on the values $M$, $g_1$.
\begin{figure}[h!]
\begin{center}
{\label{non_rot_cc_1}\includegraphics[width=5cm, angle = -90]{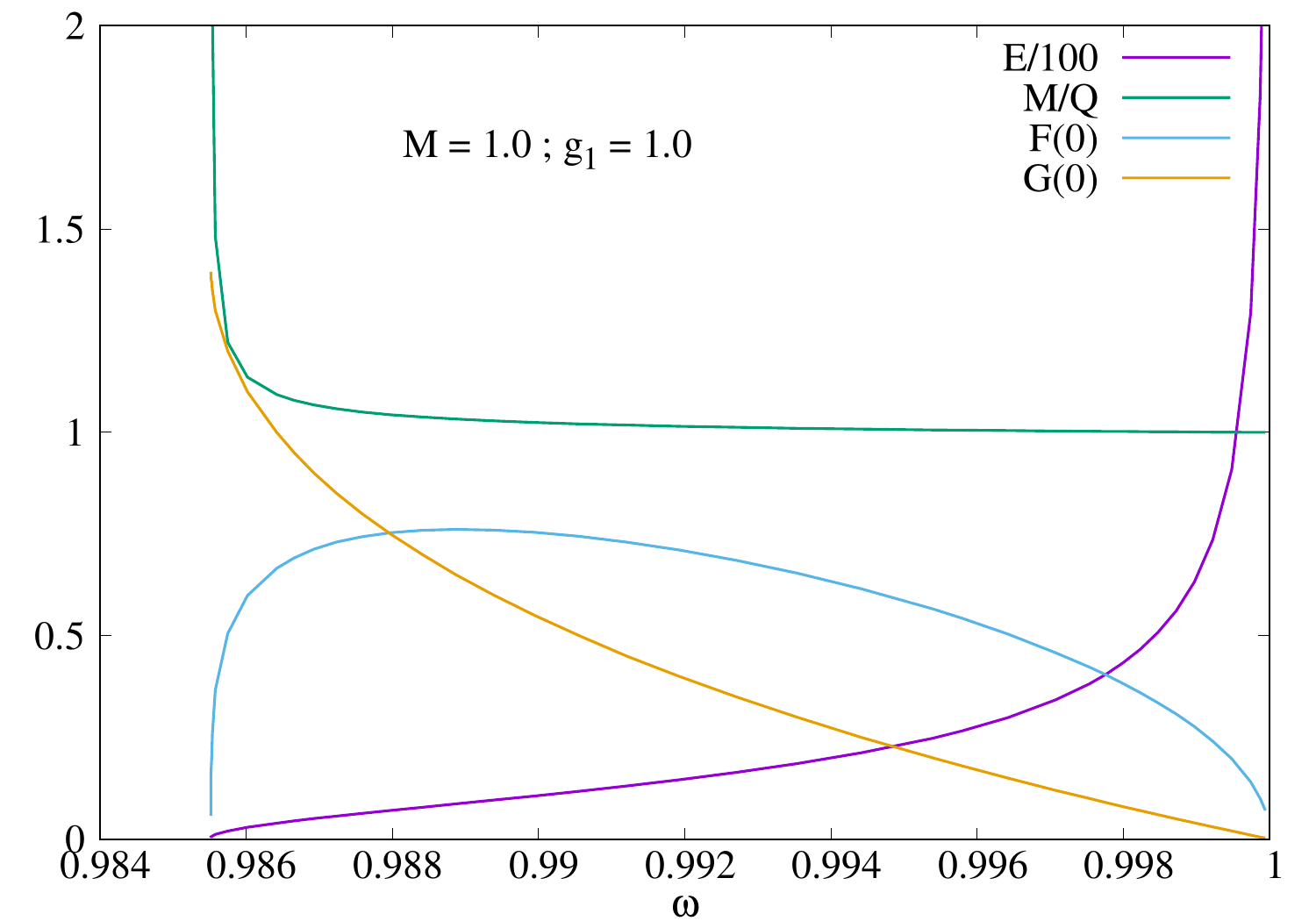}}
{\label{non_rot_cc_2}\includegraphics[width=5cm, angle = -90]{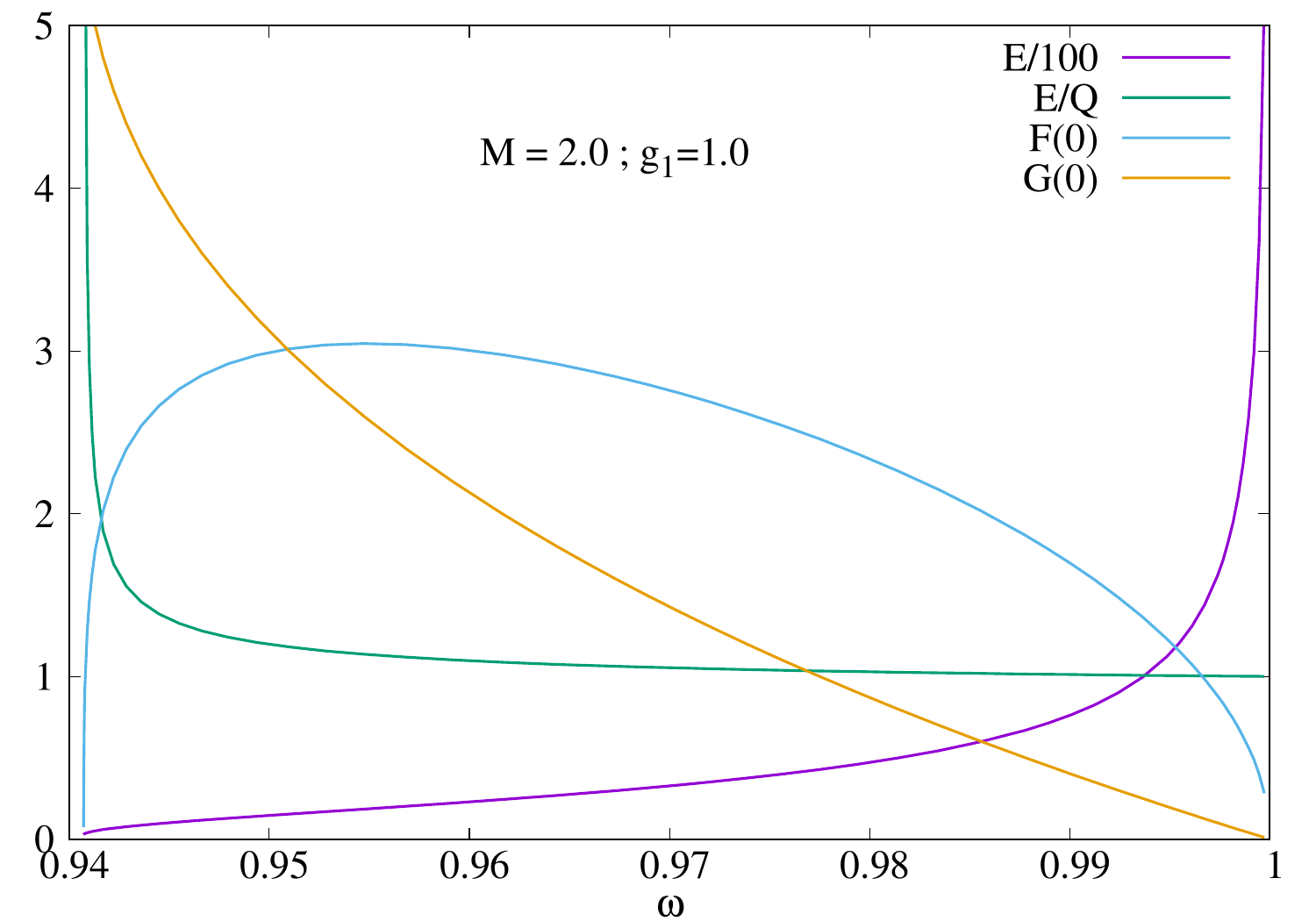}}
\end{center}
\caption{Left: The central values $F(0)$, $G(0)$, the mass and the Noether charge as function of $\omega$
for $D=3$, $M=1$, $g_1=1$
Right: the same data for $M=2$.
\label{fig1}
}
\end{figure}
Several parameters characterizing  the solutions available for the cases 
$M=1$, $g_1=1$ and $M=2$, $g_1=1$ are presented  in Fig. \ref{fig1}. The following features are observed~: 
\begin{itemize}
\item In the limit $\omega \to 1.0$, the two scalar functions $F(r),G(r)$ uniformly tend to the zero function. However these functions extend more in space and  the convergence is slow in such a way that the mass and Noether charge diverge in this limit. This limit is $\Omega\to 0$ and the vanishing of $F$ and $G$ is coherent with (\ref{profil2}).
\item In the limit $\omega \to \omega_m$ the function $F(r)$ tends uniformly to the null function. In this case
the convergence is quick enough so that the Noether charge also approaches zero. By contrast the function  $G(r) \to G_0(r) $ with $G(0) \to 5.5889$ (see above) and the mass remains finite in this limit.
\item  An increase of the parameter $M$ leads to a decrease of $\omega_m$, so to a larger interval of possible frequencies. This behaviour is in qualitative agreement with (\ref{omegam}).
\end{itemize}
Typical profiles of the functions $F(r)$, $G(r)$ and of the effective energy density $\epsilon$, defined as the integrand of (\ref{energy}), are presented in Fig. \ref{fig2} for the case 
$g_1=1$, $M = 2$. The dashed and solid lines respectively correspond
to $\omega = 0.9407$ and $\omega = 0.999$.

\begin{figure}[h!]
\begin{center}
{\label{non_rot_cc_0}\includegraphics[width=6cm, angle = -90]{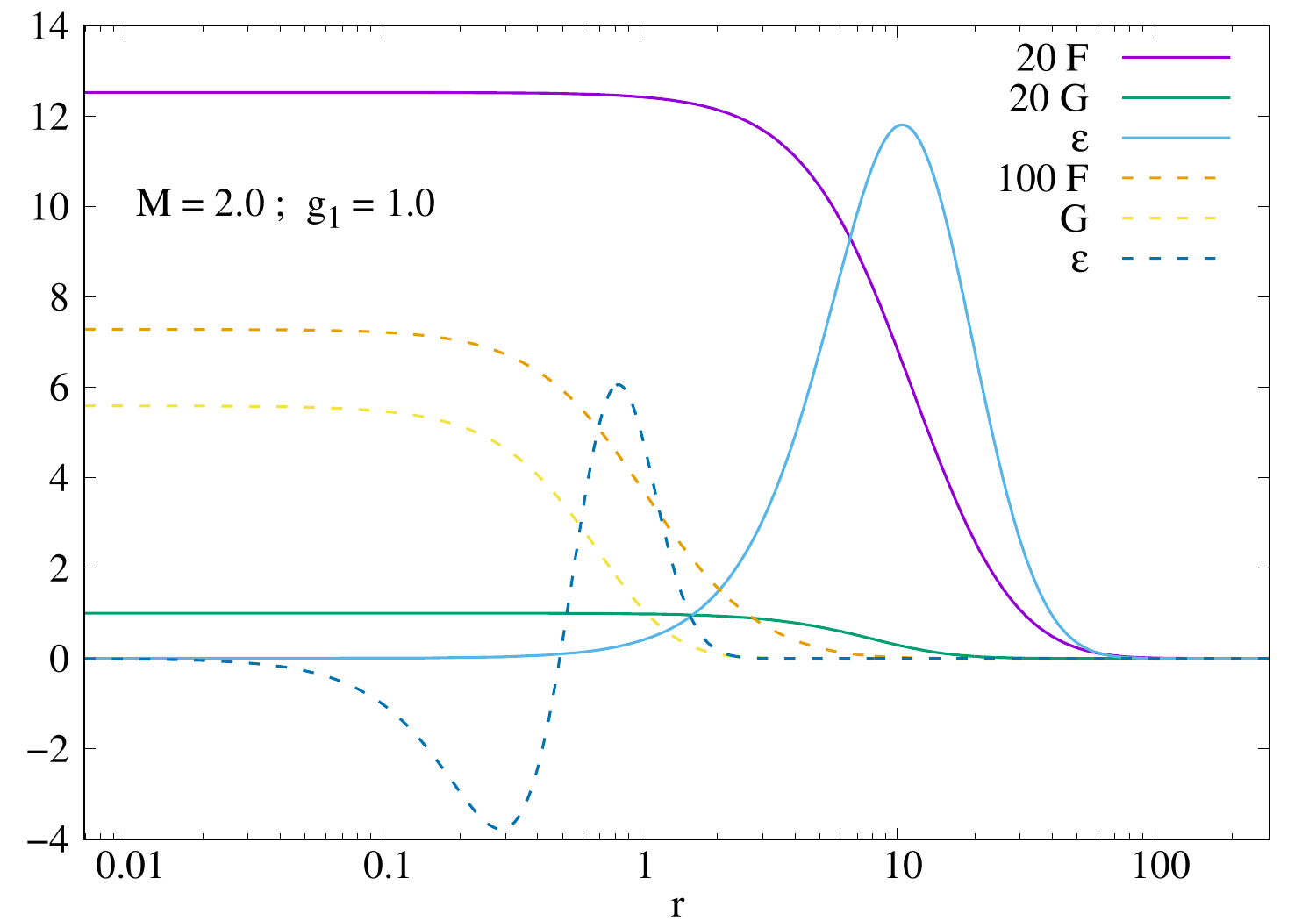}}
\end{center}
\caption{Profiles of the functions $F$, $G$ and the effective energy density $\epsilon$ for $D=3$, $M=2$, $g_1=1$
for $\omega = 0.9407$  (dashed lines) and for $\omega = 0.999$ (solid lines).
\label{fig2}
}
\end{figure}

\subsection{The case $g_1 > g_2$}
The numerical analysis reveals that the pattern of solutions observed in the case $g_1=g_2$ changes significantly when the two coupling constants are different. For this reason we find it convenient
to analyze separately the cases $g_1 > g_2$ and $g_1 < g_2$. 
The results in this section are reported for the case $M=2$.
We checked that small changes of $M$ do not affect the pattern, although the case $M \ll m$ will not be considered here.

\begin{figure}[h!]
\begin{center}
{\label{non_rot_cc_1b}\includegraphics[width=5cm, angle = -90]{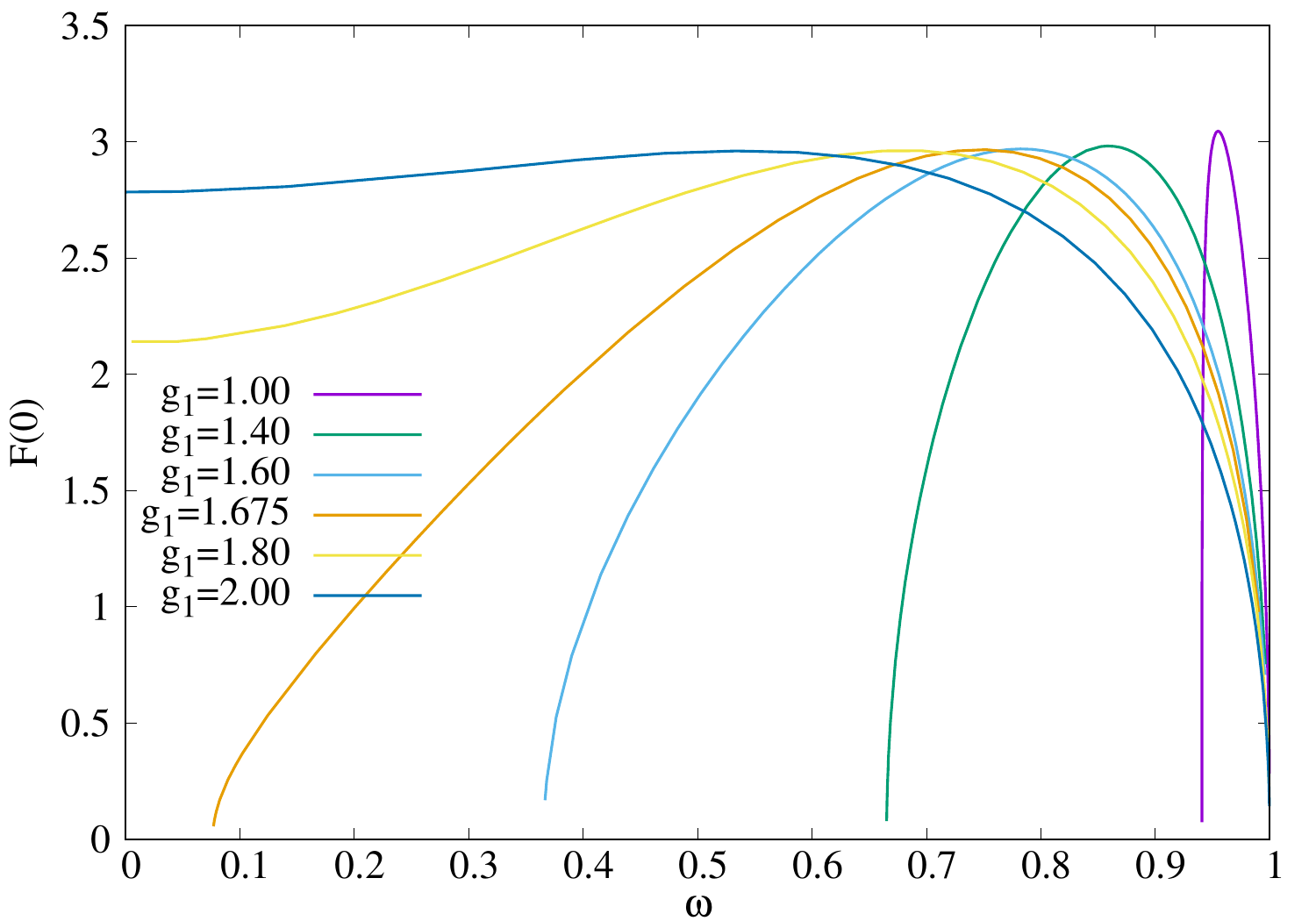}}
{\label{non_rot_cc_2b}\includegraphics[width=5cm, angle = -90]{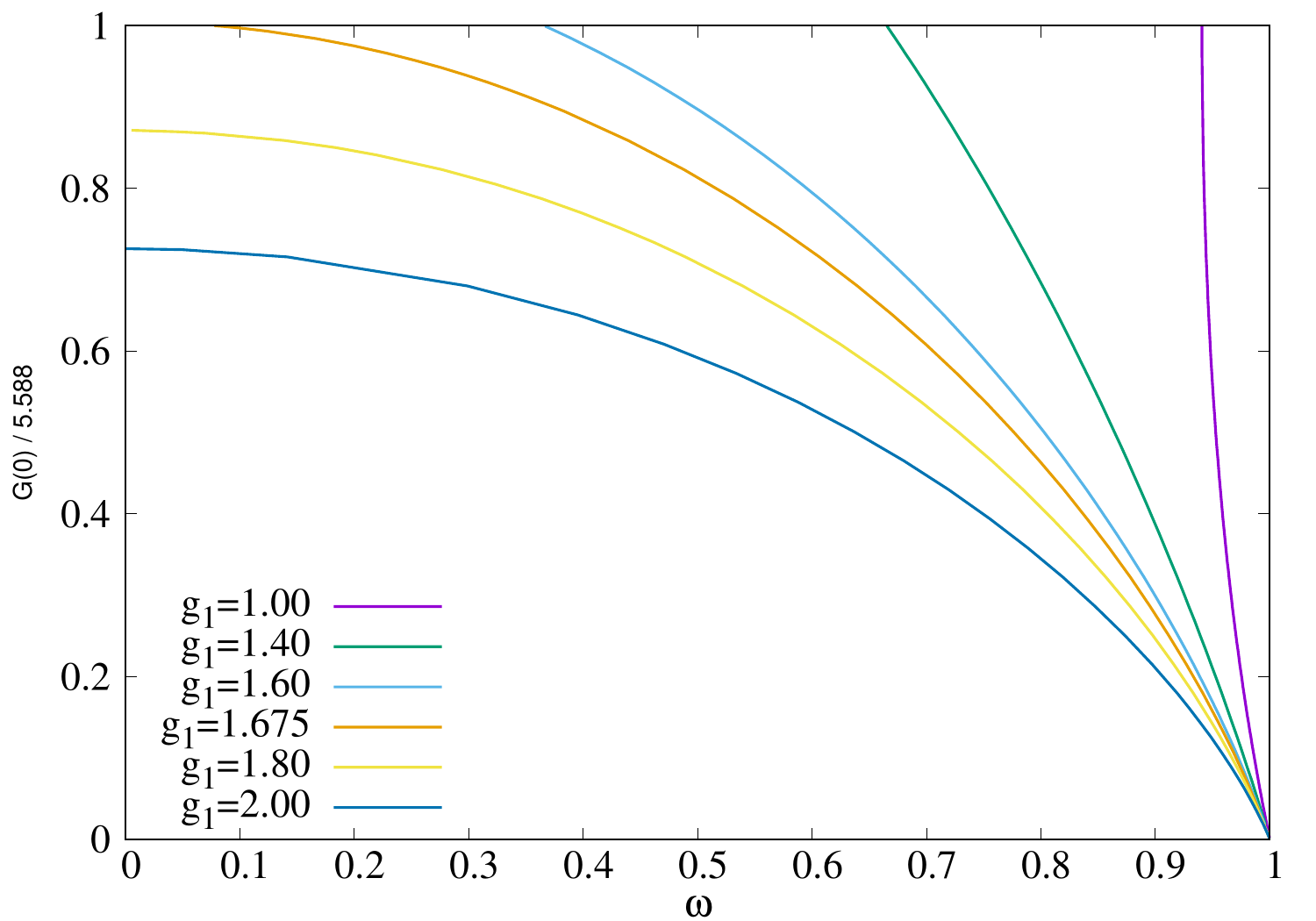}}
\end{center}
\caption{Left: The central values $F(0)$  as function of $\omega$
for $D=3, M=2$ and several values of $g_1$.
Right: the corresponding values of $G(0)$.
\label{fig_ome_fg_bis}
}
\end{figure}

\begin{figure}[h!]
\begin{center}
\includegraphics[width=6cm, angle = -90]{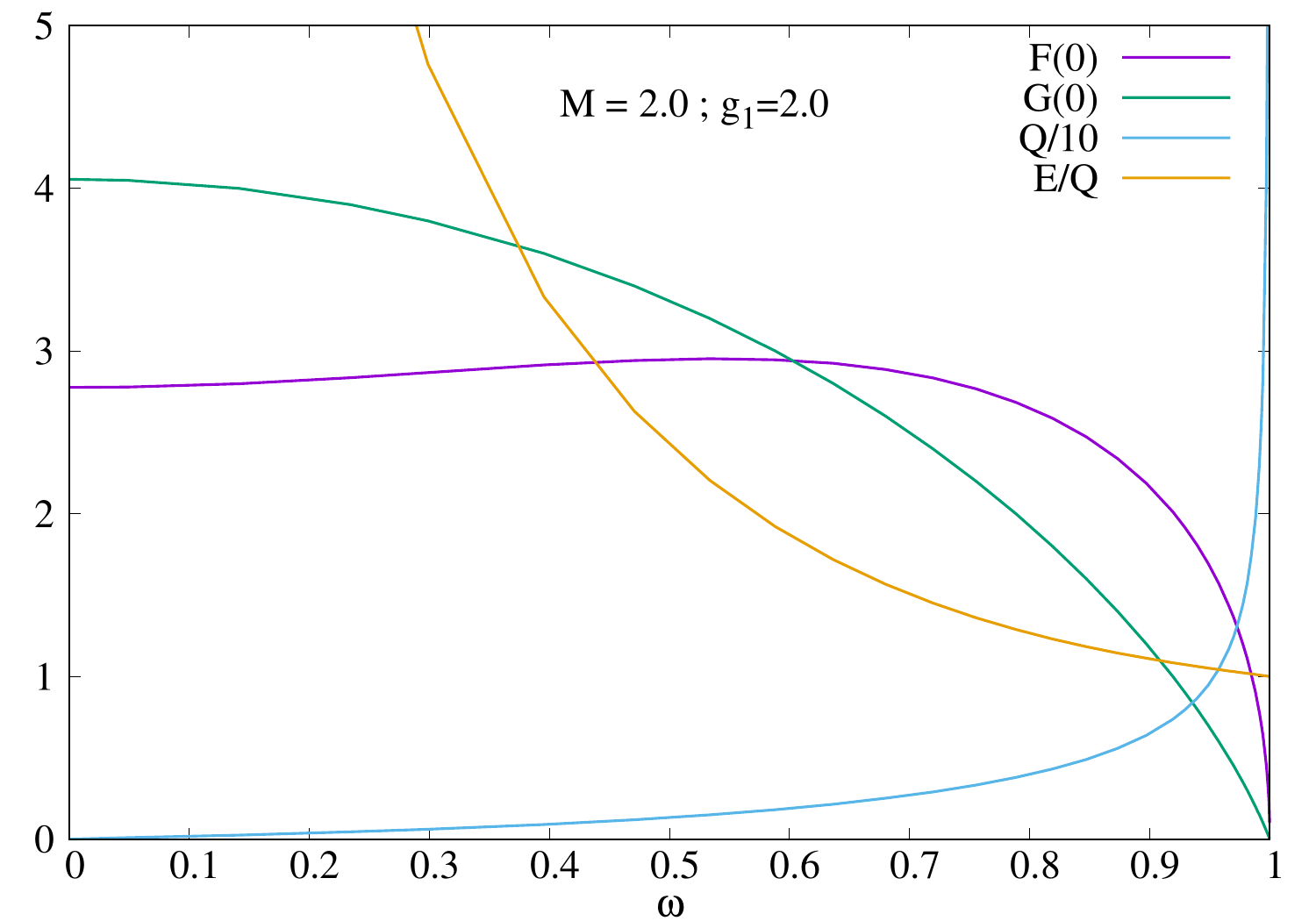}
\end{center}
\caption{Left: The central values $F(0)$, $G(0)$ and the soliton mass as function of $\omega$ for $D=3, M=2, g_1=2.0$ ~. 
\label{fig4}
}
\end{figure}

When $M=2$, and actually when $M>m$, the pattern looks similar to the case $g_1=g_2$.
Parametrizing again the solutions by the frequency $\omega$, it turns out that solutions exist for 
$\omega \in [\omega_m, 1]$ where the minimal value $\omega_m$ decreases when $g_1$ increases; this is illustrated
by Figs. \ref{fig_ome_fg_bis} where the central values 
$F(0)$, $G(0)$ are reported versus  $\omega$ for several values of $g_1$.
Interestingly, when $g_1$ is large enough, solutions exist for the full interval of frequencies $\omega \in [0,1]$.
In particular, the solutions corresponding to $\omega = 0$ are regular and real. Moreocer, it is found that real solutions exist for $g_1 \geq g_c$ with $g_c \approx 1.68$.
For $g_1 < g_c$ the branch of solutions terminates in a configuration $F(r)=0, G(r)=G_0(r)$ for $\omega \to \omega_m$,
in particular $G(0) \to 5.5889$.
Some physical parameters characterizing the case $g_1=2$ are presented in Fig. \ref{fig4}.

\subsection{The case $g_1 < g_2$} 
\begin{figure}[h!]
\begin{center}
\includegraphics[width=5cm, angle = -90]{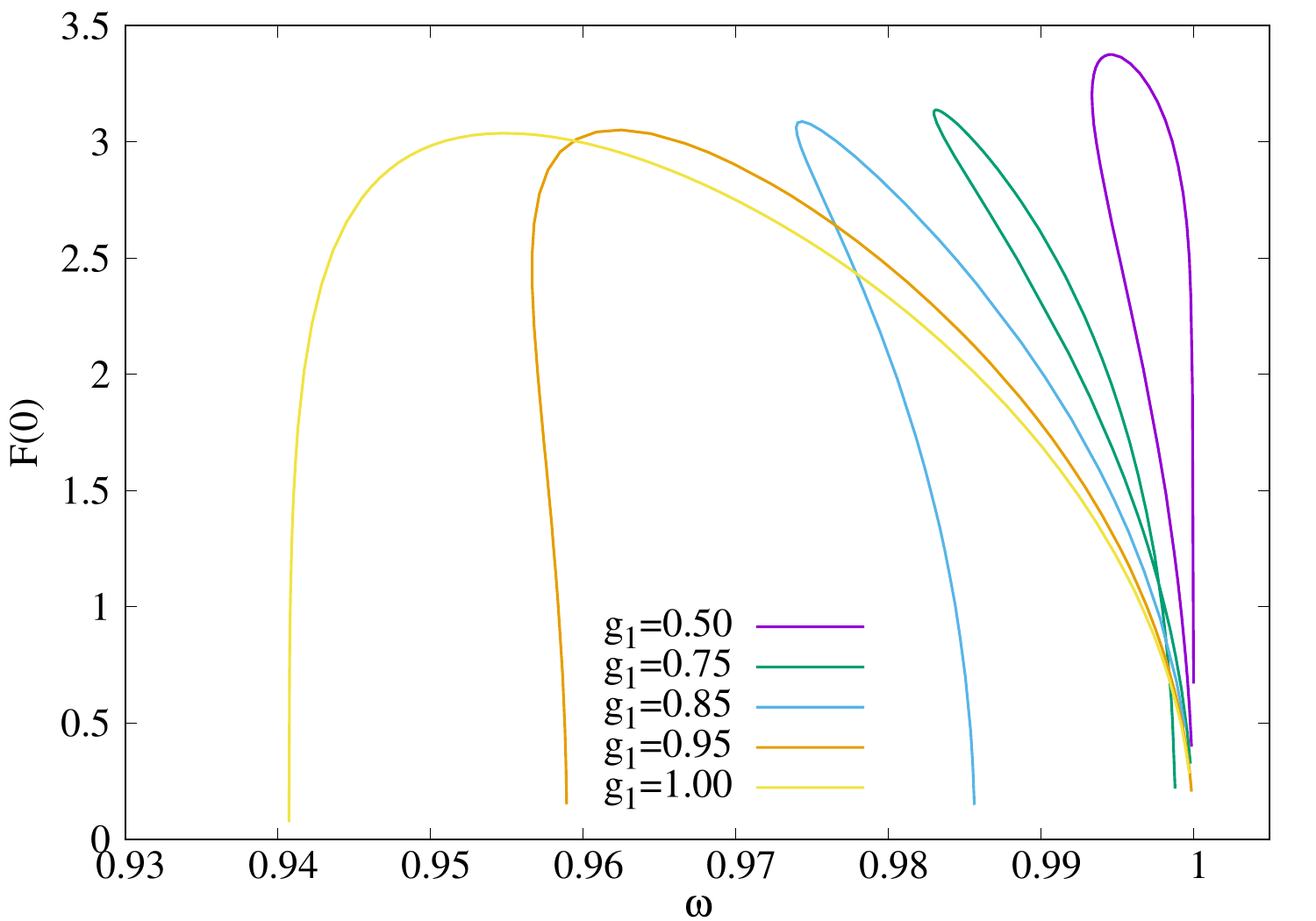}
\includegraphics[width=5cm, angle = -90]{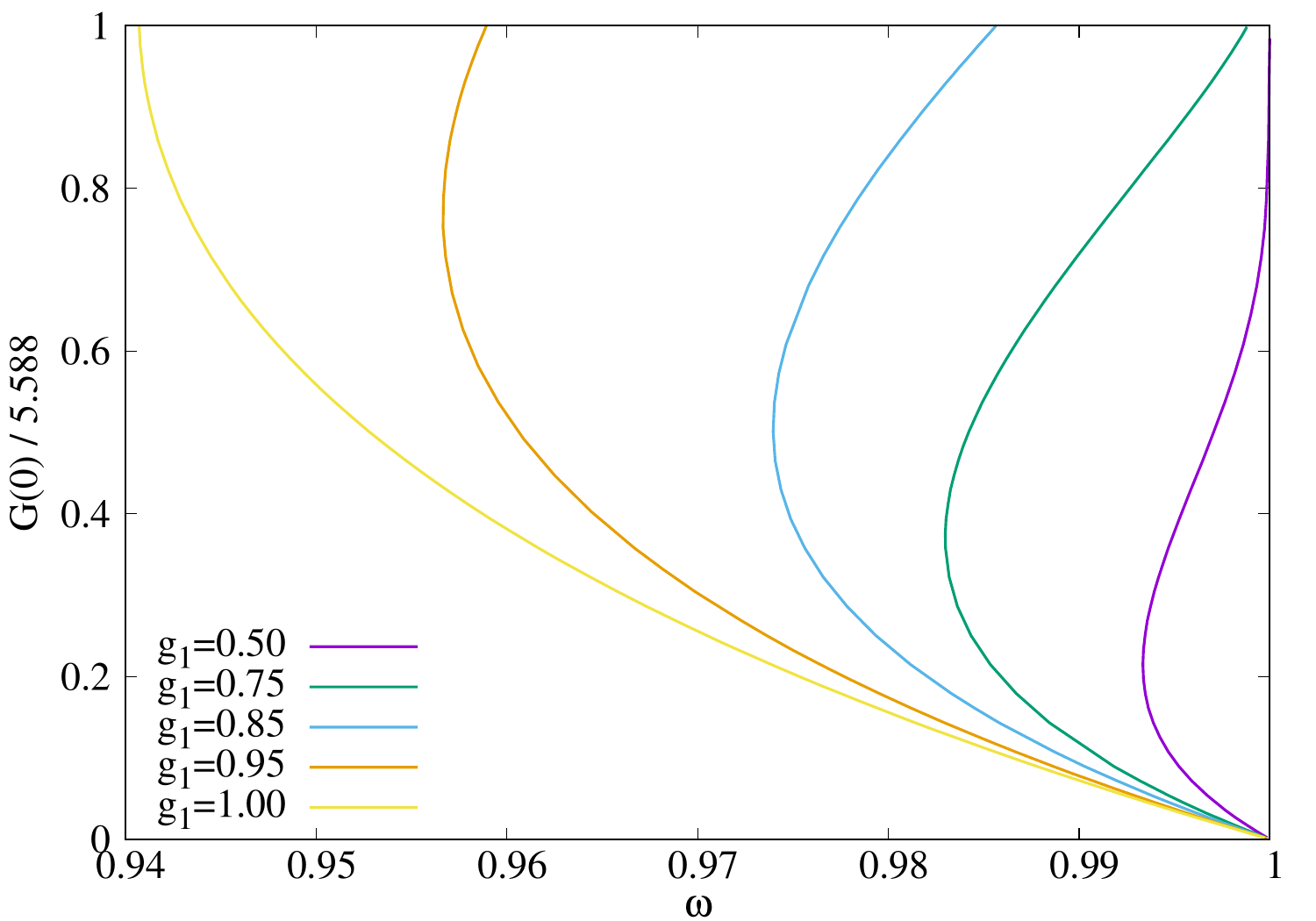}
\end{center}
\caption{Left: The central values $F(0)$  as function of $\omega$
for $D=3, M=2$ and several values of $g_1$.
Right: the corresponding values of $G(0)$~.
\label{fig_ome_fg}
}
\end{figure}

 Using again $M=2$ and the frequency $\omega$ as parameter, it is found for $g_1 < 1$ that
 families of Q-balls exist for $\omega \in [\omega_m, 1.0]$. However, a new kind of phenomenon appears
 which is illustrated by Fig. \ref{fig_ome_fg}.
It turns out that the pattern of solutions presents two branches in $\omega$ that meet into the same solution
for $\omega \to \omega_m$. In other words,
two solutions with different values of $F(0)$,  $G(0)$ and $M$ correspond to the same frequency. 

\begin{figure}[h!]
\begin{center}
\includegraphics[width=5.5cm, angle = -90]{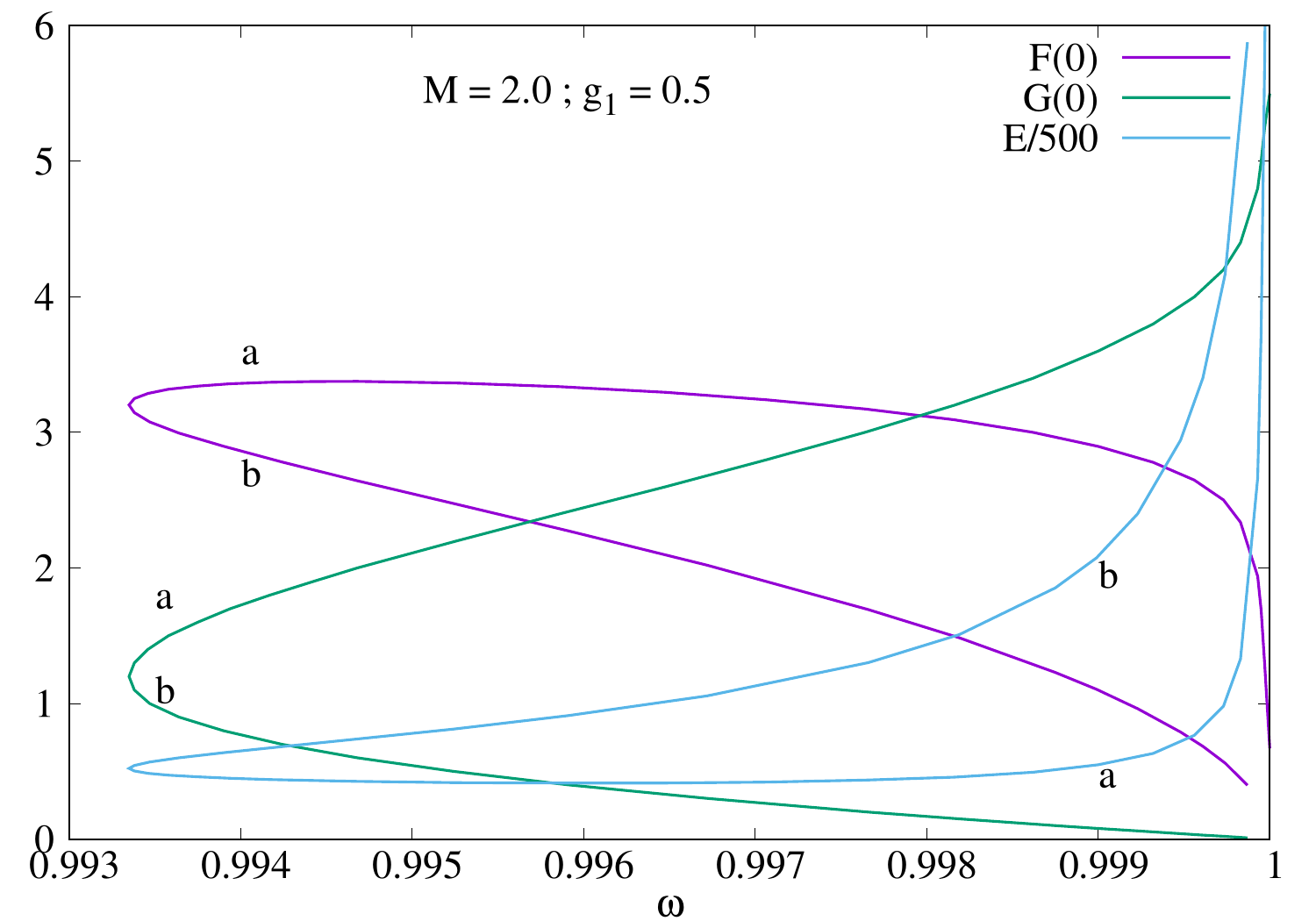}
\end{center}
\caption{Left: The central values $F(0)$, $G(0)$ and the mass  as function of $\omega$
for $D=3$, $M=2$, $g_1=0.5$ (the Noether charge is roughly equal to the mass all along). a and b label the two different branches of solutions.
\label{fig7}
}
\end{figure}

Details of the solutions corresponding to $g_1=0.5$  
are presented in Fig. \ref{fig7}. Let us finally point out that the minimal frequency $\omega_m$
increases while decreasing $g_1$.  Q-balls do likely  not exist for $g_1 < 0.35$. 

\section{Charged Q-balls}\label{sec:charged}
Among the possible extensions of our model is the promotion of the global $U(1)$ symmetry to a gauge
invariance. Here we apply the procedure to the model (\ref{lagrangian}). As usual, the gauging is achieved  by replacing the partial derivative $\partial_{\mu} \phi$ in (\ref{lagrangian}) by
a covariant derivative $D_{\mu} \phi = (\partial_{\mu} - i e A_{\mu})\phi$. The electromagnetic potential is noted $A_{\mu}$ and $e$ represents the coupling constant. A Maxwell-Faraday Lagrangian is also added. 

Completing the ansatz (\ref{ansatz}) by a spherically symmetric  electric potential $A_0 = V(r)$, $A_{i\neq0}=0$, the field equations  now read
\begin{subequations}\label{eom_em}
\begin{eqnarray}
        F'' + \frac{D-1}{r} F' &=& (m^2 - W^2)\, F - g_1 F G\ , \label{eom_em_1}\\
				G'' + \frac{D-1}{r} G' &=& M^2\, G - \frac{g_1}{2} F^2 - 3 g_2 G^2 , \label{eom_em_2} \\  
				W'' + \frac{D-1}{r} W' &=& \frac{e^2}{2} W F^2    \ ,  \label{eom_em_3} \\
				{\rm with}\quad W(r) &\equiv& \omega - e V(r) ,
\end{eqnarray}
\end{subequations}
to be solved with the boundary conditions (\ref{bound}) supplemented by $W'(0) = 0$. The quantity 
\begin{equation}\label{betadef}
	\beta \equiv \omega - e V(\infty)
\end{equation}
plays the role of $\omega$ in the last section. The electric field $V(r)$ is characterized by the chemical potential $\Phi \equiv V(r\to \infty) - V(0)$ and by the electric charge $Q_E$ such that $V(r\to \infty) \sim V(\infty) - \frac{Q_E}{r}$. The electric  charge is directly related to the Noether charge~: $Q_E = e \tilde Q_N$ by Eq. (\ref{eom_em_3}).

In our numerical study, we assume $D=3$ and  normalize $r$ and $\phi$ such that $m=1$, $g_2=1$ as previously. Along their uncharged counterparts, charged Q-balls exist in our model for $\beta \in [\beta_m, 1]$. A priori, $\beta_m$ depends on $g_1$, $M$, $e$.  We constructed several families of charged Q-balls and found, surprisingly, that 
the minimal frequency $\beta_m$ was not dependent on the electric coupling constant $e$. In \cite{Lee:1991ax} it was pointed out already that, when the electric coupling $e$ increases, the solitons have tendency to disappear for large enough coupling of the scalar field $\phi$
to the electric field. This has been recently demonstrated analytically by using thin-wall approximation in \cite{Heeck:2021zvk}. However, these last two studies couple a single field to a U(1) field. Static charged Q-balls were first investigated in the Friedberg-Lee-Sirlin model in \cite{Lee:1991bn} and the augmentation of $\beta_m$ with $e$ was numerically found in \cite{Loiko:2022noq}. To our knowledge, the existence of static charged Q-balls in a Henon-Heiles-inspired potential has never been show before, and understanding the independence of $\beta_m$ on $e$ would deserve further studies.

\begin{figure}[h!]
\begin{center}
\includegraphics[width=5cm, angle = -90]{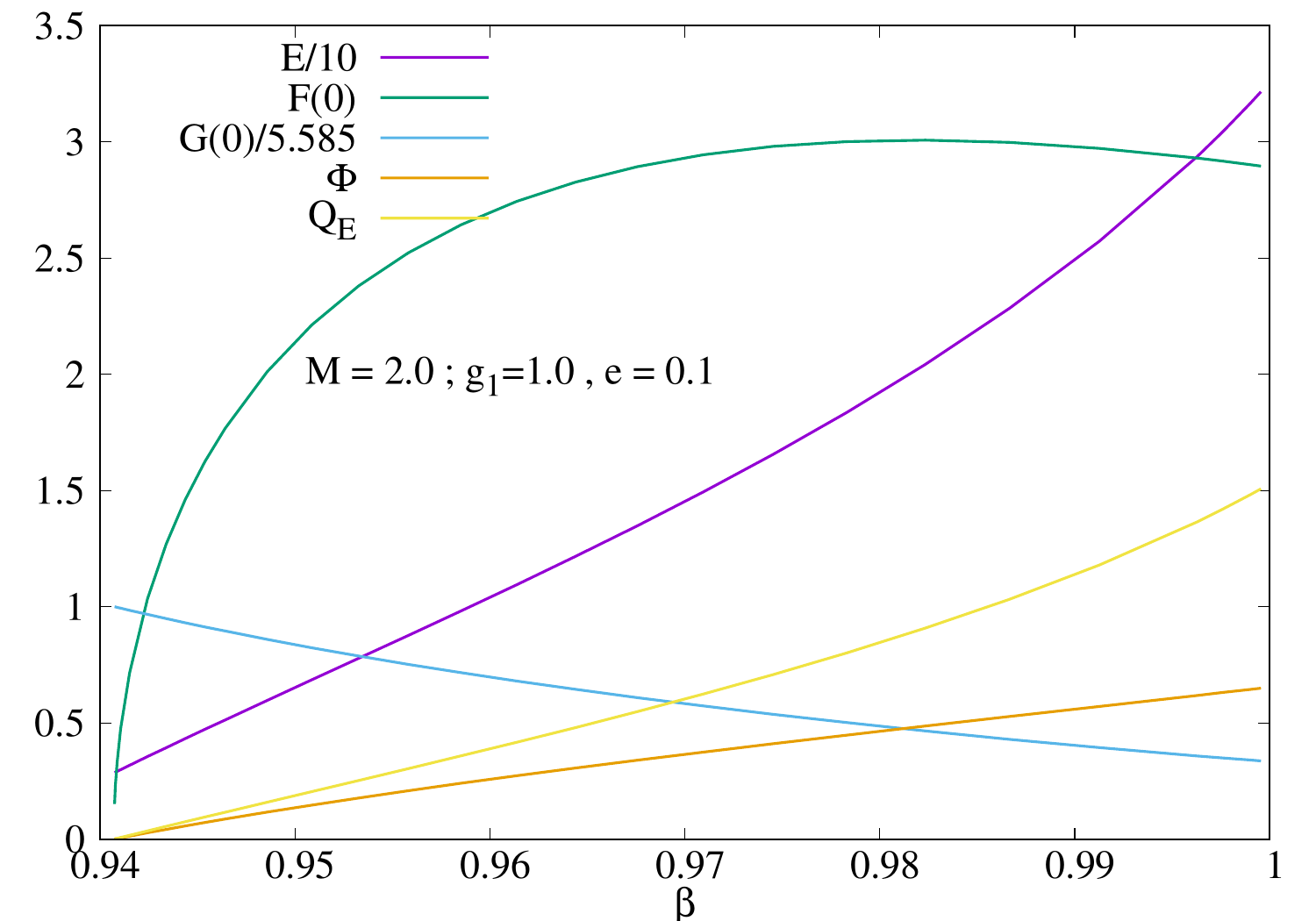}
\includegraphics[width=5cm, angle = -90]{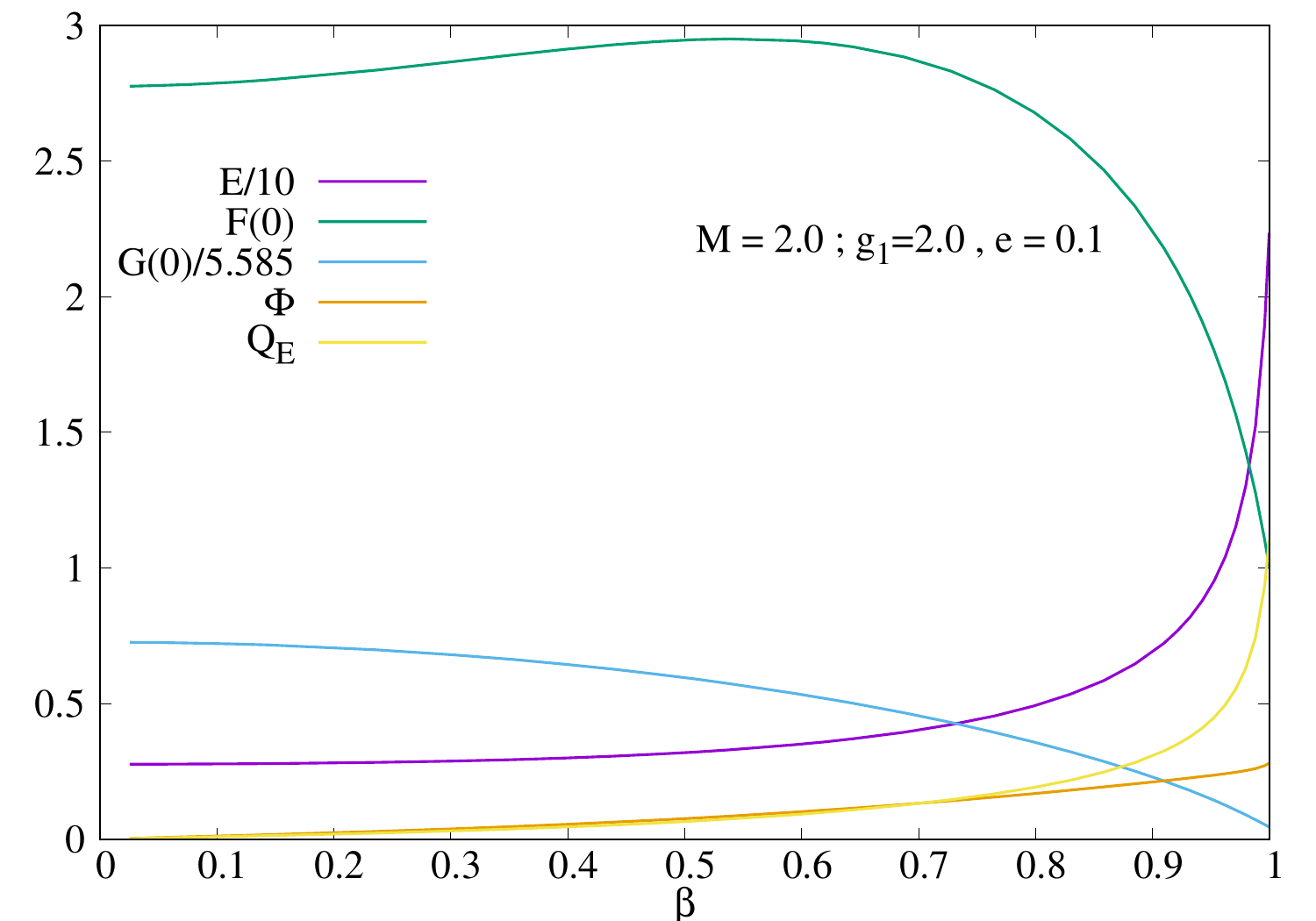}
\end{center}
\caption{Left: Some parameters  of charged Q-balls as functions of $\beta$
for $g_1=1, M=2$ and $e=0.1$ .
Right: same for $g_1=2$.
\label{fig_ome_em}
}
\end{figure}

At the approach $\beta \to \beta_m$, the pattern of solutions is similar to the uncharged case: The electric potential tends to a constant so that $\Phi$ and $Q_E$
tend to zero. The influence of the charge is more pronounced in the region $\beta \sim 1$. In this limit,
the electric parameters $\Phi$, $Q_E$ increase while the scalar fields $F(r)$, $G(r)$ do not approach the null function and remain finite. This can be understood
by the fact that, to compensate the electric repulsion, a minimal amount of -- attractive -- scalar field is necessary.
Both phenomenon are illustrated by Figs. \ref{fig_ome_em} for $M=2$, $e=0.1$ and for two values of 
the coupling constant $g_1$.

\begin{figure}[h!]
\begin{center}
\includegraphics[width=5cm, angle = -90]{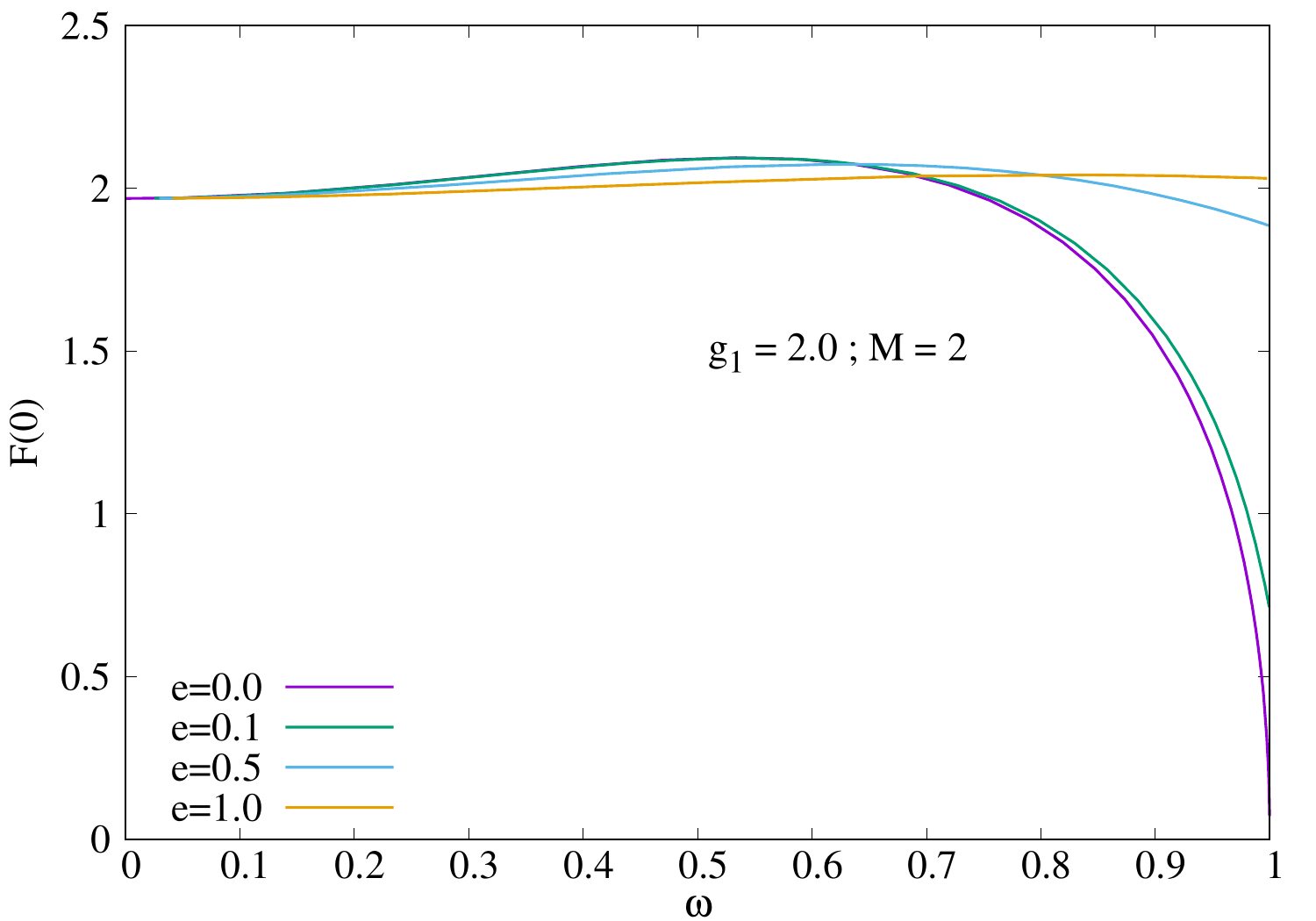}
\includegraphics[width=5cm, angle = -90]{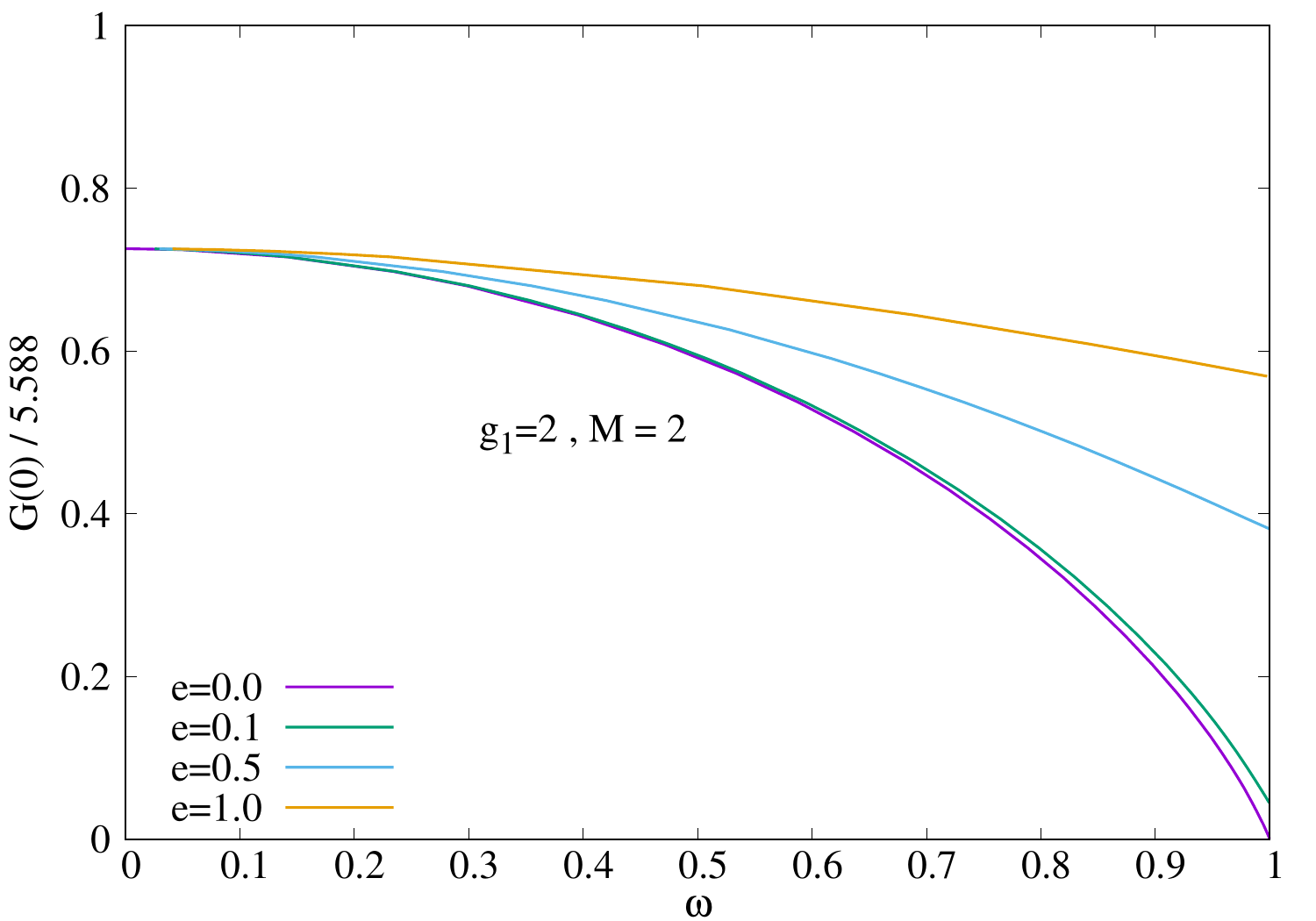}
\end{center}
\caption{Left: Central value $F(0)$ as functions of $\beta$
for $g_1=2, M=2$ and several values of $e$ .
Right: same for the value $G(0)$.
\label{fig_e_vary}
}
\end{figure}

The evolution of the solutions when increasing the parameter $e$ comes out as a natural question
which we investigated for a few values of $g_1$. When  $e$ increases, the central values  of the scalar fields $F(0)$, $G(0)$
increases and have tendency to depend weakly on $\beta$. For instance  we found that they 
 approach respectively the values $F(0)_{\beta = \beta_m}$ and $G(0)_{\beta = \beta_m}$ already for $e \sim 1$. 
This feature is illustrated by Fig. \ref{fig_e_vary}
in the case $g_1=2$. The same phenomenon holds for $g_1=1$: In this case
$F(0)_{\beta = \beta_m} \sim 0$ and $G(0)_{\beta = \beta_m} \approx 5.585$.
We limited our numerical work to $\e \leq 1$ but believe that these features  
hold for other values of $g_1$ and may persist for larger values of $e$.

\section{Summary}
We have shown that the model presented in \cite{Nugaev:2014ima}, made of one massive scalar field $\xi$ and one massive complex scalar field $\phi$ interacting via the cubic couplings $g_1 \xi \phi^{*} \phi + g_2 \xi^3$, shows Q-balls as classical solutions. In one spatial dimension, the model's effective dynamics is that of a particle in a generalized Henon-Heiles potential. We have proposed three analytical solutions of solitary-wave type. We have also shown the existence of a rich pattern of Q-balls with spherical symmetry in three spatial dimensions, which is somewhat surprising regarding the simplicity of the couplings. The Q-ball with vanishing $\phi$ has a particular status since most of the solutions we find reduce to this Q-ball when $\omega$ approaches its lower bound. When $g_1<0.35\, g_2$, no Q-ball exists. When $g_1\geq 1.68\, g_2$ however, Q-balls exist up to $\omega=0$ with nonzero values for both $\phi$ and $\xi$. 
Finally, we showed that solutions continue to exist when a coupling to an electric field is supplemented
and we have discussed how their domain of existence evolves in response to the electric coupling constant.
We leave the exploration of solutions with rotation or nodes, or of boson-star type, for future works.


\begin{thebibliography}{10}
	
	\bibitem{Coleman:1985ki}
	Sidney~R. Coleman.
	\newblock {Q-balls}.
	\newblock {\em Nucl. Phys. B}, 262(2):263, 1985.
	\newblock [Addendum: Nucl.Phys.B 269, 744 (1986)].
	
	\bibitem{Lee:1991ax}
	T.~D. Lee and Y.~Pang.
	\newblock {Nontopological solitons}.
	\newblock {\em Phys. Rept.}, 221:251--350, 1992.
	
	\bibitem{Volkov:2002aj}
	Mikhail~S. Volkov and Erik Wohnert.
	\newblock {Spinning Q balls}.
	\newblock {\em Phys. Rev.}, D66:085003, 2002.
	
	\bibitem{charged_qb}
	Kimyeong Lee, Jaime~A. Stein-Schabes, Richard Watkins, and Lawrence~M. Widrow.
	\newblock Gauged q balls.
	\newblock {\em Phys. Rev. D}, 39:1665--1673, Mar 1989.
	
	\bibitem{Brihaye:2007tn}
	Yves Brihaye and Betti Hartmann.
	\newblock Interacting q-balls.
	\newblock {\em Nonlinearity}, 21(8):1937, jul 2008.
	
	\bibitem{sirlin}
	R.~Friedberg, T.~D. Lee, and A.~Sirlin.
	\newblock Class of scalar-field soliton solutions in three space dimensions.
	\newblock {\em Phys. Rev. D}, 13:2739--2761, May 1976.
	
	\bibitem{Loiko:2019gwk}
	V.~Loiko and Ya. Shnir.
	\newblock Q-balls in the u(1) gauged friedberg-lee-sirlin model.
	\newblock {\em Physics Letters B}, 797:134810, 2019.
	
	\bibitem{Nugaev:2014ima}
	E.Ya. Nugaev.
	\newblock Hénon–heiles potential as a bridge between nontopological solitons
	of different types.
	\newblock {\em Communications in Nonlinear Science and Numerical Simulation},
	20(2):443--446, 2015.
	
	\bibitem{Nugaev:2019vru}
	E.~Ya. Nugaev and A.V. Shkerin.
	\newblock {Review of Nontopological Solitons in Theories with $U(1)$-Symmetry}.
	\newblock {\em J. Exp. Theor. Phys.}, 130(2):301--320, 2020.
	
	\bibitem{hh}
	Michel {Henon} and Carl {Heiles}.
	\newblock {The applicability of the third integral of motion: Some numerical
		experiments}.
	\newblock {\em Astron. J.}, 69:73, February 1964.
	
	\bibitem{colsys}
	U.~Ascher, J.~Christiansen, and R.~D. Russell.
	\newblock {A Collocation Solver for Mixed Order Systems of Boundary Value
		Problems}.
	\newblock {\em Math. Comput.}, 33(146):659--679, 1979.
	
	\bibitem{10.3389/fspas.2022.945236}
	Sawsan Alhowaity, Elbaz~I. Abouelmagd, Zouhair Diab, and Juan L.~G. Guirao.
	\newblock Calculating periodic orbits of the hénon–heiles system.
	\newblock {\em Frontiers in Astronomy and Space Sciences}, 9, 2023.
	
	\bibitem{zlotos}
	Euaggelos~E. Zotos.
	\newblock Classifying orbits in the classical h{\'e}non--heiles hamiltonian
	system.
	\newblock {\em Nonlinear Dynamics}, 79(3):1665--1677, 2015.
	
	\bibitem{verhoeven}
	C.~Verhoeven, M.~Musette, and R.~Conte.
	\newblock {Integration of a generalized Hénon–Heiles Hamiltonian}.
	\newblock {\em Journal of Mathematical Physics}, 43(4):1906--1915, 04 2002.
	
	\bibitem{WOJCIECHOWSKI1984277}
	Stefan Wojciechowski.
	\newblock Separability of an integrable case of the henon-heiles system.
	\newblock {\em Physics Letters A}, 100(6):277--278, 1984.
	
	\bibitem{conte}
	R~Conte and M~Musette.
	\newblock Link between solitary waves and projective riccati equations.
	\newblock {\em Journal of Physics A: Mathematical and General}, 25(21):5609,
	nov 1992.
	
	\bibitem{Heeck:2021zvk}
	Julian Heeck, Arvind Rajaraman, Rebecca Riley, and Christopher~B. Verhaaren.
	\newblock {Mapping Gauged Q-Balls}.
	\newblock {\em Phys. Rev. D}, 103(11):116004, 2021.
	
	\bibitem{Lee:1991bn}
	Chul~H. Lee and Seung~Un Yoon.
	\newblock {Existence and stability of gauged nontopological solitons}.
	\newblock {\em Mod. Phys. Lett. A}, 6:1479--1486, 1991.
	
	\bibitem{Loiko:2022noq}
	Victor Loiko and Yakov Shnir.
	\newblock {Q-ball stress stability criterion in U(1) gauged scalar theories}.
	\newblock {\em Phys. Rev. D}, 106(4):045021, 2022.
	
\end{thebibliography}

\end{document}